# Scientific Evidence for "Hydrostatic Shock"

Michael Courtney, PhD
Ballistics Testing Group, P.O. Box 24, West Point, NY 10996
Michael_Courtney@alum.mit.edu

Amy Courtney, PhD
Department of Physics, United States Military Academy, West Point, NY 10996
Amy_Courtney@post.harvard.edu

**Abstract:**
This paper reviews the scientific support for a ballistic pressure wave radiating outward from a penetrating projectile and causing injury and incapacitation. This phenomenon is known colloquially as "hydrostatic shock." The idea apparently originates with Col. Frank Chamberlin, a World War II trauma surgeon and wound ballistics researcher. The paper reviews claims that hydrostatic shock is a myth and considers supporting evidence through parallels with blast, describing the physics of the pressure wave, evidence for remote cerebral effects, and remote effects in the spine and other internal organs. Finally, the review considers the levels of energy transfer required for the phenomenon to be readily observed.

Debates between bullets that are "light and fast" vs. "slow and heavy" often refer to "hydrostatic shock," which describes remote wounding and incapacitating effects in living targets in addition to tissue crushed by direct bullet impact. Considerable evidence shows that "hydrostatic shock" can produce remote neural damage and rapid incapacitation.

**Background**
It is unclear when "hydrostatic shock" was first used to describe bullet effects, but Frank Chamberlin, a World War II trauma surgeon and ballistics researcher, noted remote pressure wave effects. Col. Chamberlin described "explosive effects" and "hydraulic reaction" of bullets in tissue:

*. . . liquids are put in motion by 'shock waves' or hydraulic effects . . . with liquid filled tissues, the effects and destruction of tissues extend in all directions far beyond the wound axis.* [1]

He avoided the ambiguous use of the term "shock" because it can refer to either a specific kind of pressure wave associated with explosions and supersonic projectiles or to a medical condition in the body.

Col. Chamberlin recognized that many theories have been advanced in wound ballistics. During World War II, he commanded an 8500 bed hospital center that treated over 67,000 patients during the fourteen months that he operated it. P.O. Ackley estimates that 85% of the patients were suffering from gun shot wounds.[2] Col. Chamberlin spent many hours interviewing patients as to their reactions to bullet wounds. He also conducted many live animal experiments after his tour of duty. On the subject of wound ballistics theories, he wrote:

*If I had to pick one of these theories as gospel, I'd still go along with the Hydraulic Reaction of the Body Fluids plus the reactions on the Central Nervous System.*[1]

Other World War II era scientists noted remote pressure wave effects in the peripheral nerves.[3][4] There was support for the idea of remote neural effects of ballistic pressure waves in the medical and scientific communities, but the term "hydrostatic shock" and similar phrases including "shock" were used mainly by gunwriters (such as Jack O'Conner[5]) and the small arms industry (such as Roy Weatherby.[6])

**A Myth?**
Dr. Martin Fackler, a Vietnam-era trauma surgeon and ballistics researcher, claimed that hydrostatic shock had been disproven. Specifically, he said the assertion that a pressure wave plays a role in injury or incapacitation is a myth.[7] Others, including ballistics experts with the FBI, expressed similar views.[8][9]

In support of his claim, Dr. Fackler argued that a lithotriptor (a medical device used to break up kidney stones with sonic pressure waves) produces no damage to soft tissues. Since a lithotriptor produces pressure waves larger than those caused by most handgun bullets, he concluded that ballistic pressure waves cannot damage tissue either.[11] However, Fackler's claim by analogy has been disproven. Tissue damage due to lithotriptors has been widely documented.[12][13][14]

Other than analogies to lithotriptors and a 1947 study which did not examine neural tissue,[10] authors arguing against ballistic effects remote from the wound channel have not provided experimental evidence. Because subtle damage in neural tissues was difficult to detect, denials persisted for some time. However, as discussed below, scientific progress eventually afforded considerable support for it.

**Parallels with Blast (Explosions)**
A shock wave can be created when fluid is rapidly displaced by an explosive or projectile. Duncan MacPherson, a member of the defunct International Wound Ballistics Association and author of the book, Bullet Penetration, claimed that shock waves cannot result from bullet impacts with tissue.[9] In contrast, Brad Sturtevant, a leading researcher in shock wave physics at Caltech for many decades, found that shock waves can result from handgun bullet impacts in tissue.[15] Other sources also indicate that ballistic impacts can create shock waves in tissue.[16][17][18]

Blast and ballistic pressure waves have physical similarities. They also have similarities in how they cause neural effects in the brain. In tissue, both types of pressure waves have similar magnitudes, duration, and frequency characteristics. Both have been shown to cause damage in the area of the brain known as the hippocampus.[19][20][21] It has been hypothesized that both can reach the brain from the thoracic cavity via major blood vessels.

For example, Ibolja Cernak, a leading researcher in blast wave injury at the Applied Physics Laboratory at Johns Hopkins University, hypothesized, "alterations in brain function following blast exposure are induced by kinetic energy transfer of blast overpressure via great blood vessels in abdomen and thorax to the central nervous system."[22] This hypothesis is supported by observations of neural effects in the brain from localized blast exposure focused on the lungs in animal experiments.[20]

"Hydrostatic shock" expresses the idea that organs can be damaged by the pressure wave independently from direct contact with the penetrating projectile. If one interprets the "shock" in "hydrostatic shock" to refer to physiological effects rather than physical wave characteristics, the question of whether the pressure waves satisfy the definition of "shock wave" is unimportant. There is compelling scientific evidence supporting the ability of a ballistic pressure wave to create tissue damage and incapacitation in living targets.

**Physics of Ballistic Pressure Waves**
A number of investigators have studied the physics of ballistic pressure waves created when a ballistic projectile enters a viscous medium.[23][24][25] Ballistic impacts produce pressure waves that propagate near the speed of sound.

Lee et al. present an analytical model showing that unreflected ballistic pressure waves are well approximated by an exponential decay, which is similar to blast pressure waves.[23] Lee et al. also note the importance of energy transfer, writing, "an accurate estimation of the kinetic energy loss by a projectile is always important in determining the ballistic waves."

The rigorous methods of Lee et al. require knowing the drag coefficient and frontal area of the penetrating projectile at every instant of the penetration. Since this is not generally possible with expanding handgun bullets, a model for estimating the peak pressure waves of handgun bullets from the impact energy and penetration depth in ballistic gelatin has been developed.[26] This model agrees with the more rigorous approach of Lee et al. in cases where they can both be applied. For expanding handgun bullets, the peak pressure wave magnitude is proportional to the bullet's kinetic energy divided by the penetration depth.

**Remote Cerebral Effects of Ballistic Pressure Waves**
A Swedish research group (Goransson et al.) were the first contemporary researchers to present compelling evidence for remote cerebral effects from bullet impact to an extremity.[27] They observed significantly reduced electrical activity in the brain via EEG readings from pigs shot in the thigh. Investigating further, another research group (Suneson et al.) implanted high-speed pressure transducers into the brain of pigs and demonstrated that a significant pressure wave reaches the brain of pigs shot in the thigh.[18][28] These scientists observed breathing disruption, depressed EEG readings, and neural damage in the brain caused by the distant effects of the ballistic pressure wave originating in the thigh.

These results were later confirmed and expanded upon by a Chinese research group (Wang et al.) conducting an experiment in dogs[19] which confirmed that distant effect exists in the central nervous system after a missile impact to an extremity. "A high-frequency oscillating pressure wave with large amplitude and short duration was found in the brain. . ." They observed significant damage in both the hypothalamus and hippocampus regions of the brain due to remote effects of the ballistic pressure wave.

**Remote Pressure Wave Effects in the Spine and Internal Organs**
The brain is not the only organ subject to remote pressure wave effects. In a study of handgun injury, Sturtevant found that pressure waves from a bullet impact in the torso can reach the spine. Moreover, a focusing effect from concave surfaces can concentrate the pressure wave on the spinal cord, producing significant injury.[15] This is consistent with other work showing remote spinal cord injuries from ballistic impacts.[38][39]

A group at Johns Hopkins University (Roberts et al.) has published both experimental work and finite element modeling showing considerable pressure wave magnitudes in the thoracic cavity produced by handgun projectiles stopped by a Kevlar vest.[16][17] For example, an 8 gram projectile at 360 m/s impacting a NIJ level II vest over the sternum can produce an estimated pressure wave level of nearly 2.0 MPa (300 PSI) in the heart and of nearly 1.5 MPa (220 PSI) in the lungs. Impacting over the liver can produce an estimated pressure wave level of 2.0 MPa (300 PSI) in the liver.

**Energy Transfer Required for Remote Neural Effects**

Our own research (Courtney and Courtney) supports the conclusion that handgun levels of energy transfer can produce pressure waves leading to incapacitation and injury.[29][30][26][31][32] The work of Suneson et al. also suggests that remote neural effects can occur with levels of energy transfer possible with handguns (roughly 500 ft-lbs/700 joules).

Using sensitive biochemical techniques, the work of Wang et al. suggests even lower impact energy thresholds for remote neural injury to the brain. In analysis of experiments of dogs shot in the thigh they report highly significant neural effects in the hypothalamus and hippocampus (regions of the brain) with energy transfer levels close to 150 ft-lbs. They also report less significant remote neural effects in the hypothalamus with energy transfer just under 100 ft-lbs.[19]

Even though Wang et al. document remote neural damage for low levels of energy transfer, these levels of neural damage are probably too small to contribute to rapid incapacitation. Courtney and Courtney suggest that remote neural effects only begin to make significant contributions to rapid incapacitation for ballistic pressure wave levels above 500 PSI (corresponds to transferring roughly 300 ft-lbs in 12 inches of penetration) and become easily observable above 1000 PSI (corresponds to transferring roughly 600 ft-lbs in 1 foot of penetration).[29] Incapacitating effects in this range of energy transfer are consistent with observations of remote spinal injuries,[15] observations of suppressed EEGs and breathing interruptions in pigs,[27][33] and with observations of incapacitating effects of ballistic pressure waves without a wound channel.[34]

**Other Scientific Findings**

The scientific literature contains other findings regarding injury mechanisms of ballistic pressure waves. Ming et al. report that ballistic pressure waves can break bones.[35] Tikka et al. reports abdominal pressure changes produced in pigs hit in one thigh.[36] Akimov et al. report on injuries to the nerve trunk from gunshot wounds to the extremities.[37]

**Recommendations**

The FBI recommends that loads intended for self-defense and law enforcement applications meet a minimum penetration requirement of 12" in ballistic gelatin.[8] Maximizing ballistic pressure wave effects requires transferring maximum energy in a penetration distance that meets this requirement. In addition, bullets that fragment and meet minimum penetration requirements generate higher pressure waves than bullets which do not fragment. Understanding the potential benefits of remote ballistic pressure wave effects leads us to favor loads with at least 500 ft-lbs of energy.

With a handgun, no wounding mechanism can be relied on to produce incapacitation 100% of the time within the short span of most gunfights. Selecting a good self-defense load is only a small part of surviving a gunfight. You have to hit an attacker to hurt him, and you need a good plan for surviving until your hits take effect. Get good training, practice regularly, learn to use cover, and pray that you will never have a lethal force encounter armed only with a handgun.

## About the Authors

*Amy Courtney* serves on the Physics faculty of the United States Military Academy at West Point. She earned a MS in Biomedical Engineering from Harvard University and a PhD in Medical Engineering and Medical Physics from a joint Harvard/MIT program. She has taught Anatomy and Physiology as well as Physics. She has served as a research scientist at the Cleveland Clinic.

*Michael Courtney* earned a PhD in experimental Physics from the Massachusetts Institute of Technology. He has served as the Director of the Forensic Science Program at Western Carolina University and also been a Physics Professor, teaching Physics, Statistics, and Forensic Science.